\documentclass{svproc}
\usepackage{url}
\usepackage[ruled]{algorithm2e}
\usepackage{graphicx}

\begin{document}
\mainmatter              
\title{AMALGAM: A Matching Approach to fairfy tabuLar data with  knowledGe grAph Model}
\titlerunning{AMALGAM}  
%
\author{}
\institute{}
\author{Rabia Azzi\inst{} \and Gayo Diallo\inst{}}
\authorrunning{Rabia Azzi et al.} 
%
\tocauthor{Rabia Azzi and Gayo Diallo}
\institute{BPH Center/INSERM U1219, Univ. Bordeaux, 
           F-33000, France,\\
\email{first.last@u-bordeaux.fr}}

\maketitle              

\begin{abstract}
In this paper we present \texttt{AMALGAM}, a matching approach to fairify tabular data with the use of a knowledge graph. The ultimate goal is to provide fast and efficient approach to annotate tabular data with entities from a background knowledge. The approach combines lookup and filtering services combined with text pre-processing techniques. Experiments conducted in the context of the 2020 Semantic Web Challenge on Tabular Data to Knowledge Graph Matching with both Column Type Annotation and Cell Type Annotation tasks showed promising results.
\keywords{Tabular Data, Knowledge Graph, Entity Linking}
\end{abstract}

\section{Introduction}
Making web data complying with the FAIR\footnote{FAIR stands for: Findable, Accessible, Interoperable and Reusable} principles has become a necessity in order to facilitate their discovery and reuse \cite{WilkinsonFAIR}. The value for the knowledge discovery of implementing FAIR is to increase, data integration, data cleaning, data mining, machine learning and knowledge discovery tasks. Successfully implemented FAIR principles will improve the value of data by making them findable, accessible and resolve semantic ambiguities. Good data management is not a goal in itself, but rather is the key conduit leading to knowledge discovery and acquisition, and to subsequent data and knowledge integration and reuse by the community after the data publication process \cite{ref_Mark2016}. 

Semantic annotation could be considered as a particular knowledge acquisition task \cite{Diallo2006,Drame2016,ref_Handschuh2007}. The semantic annotation process may rely on formal metadata resources described with an Ontology, even sometimes with multiple ontologies thanks to the use of semantic repositories \cite{Diallo2011}.
Over the last years, tables are one of the most used formats to share results and data. In this field, a set of systems for matching web tables to knowledge bases has been developed \cite{ref_lncs1,ref_article2}. They can be categorized in two main tasks: structure and semantic annotation. The structure annotation deals with tasks such as data type prediction and table header annotation \cite{ref_article3}. Semantic annotation involves matching table elements into KG \cite{KG} e.g., columns to class and cells to entities \cite{ref_lncs2,ref_lncs3}.

Recent years have seen an increasing number of works on Semantic Table Interpretation. In this context, SemTab 2020\footnote{http://www.cs.ox.ac.uk/isg/challenges/sem-tab/2020/index.html} has emerged as an initiative which aims at benchmarking systems which deals with annotating tabular data with entities from a KG, referred as table annotation \cite{ref_semtab}. SemTab is organised into three tasks, each one with several evaluation rounds. For the 2020 edition for instance, it involves: (i) assigning a semantic type (e.g., a KG class) to a column (\texttt{CTA}); (ii) matching a cell to a KG entity (\texttt{CEA}); (iii) assigning a KG property to the relationship between two columns (\texttt{CPA}).

Our goal is to automatically annotate on the fly tabular data. Thus, our annotation approach is fully automated, as it does not need prior information regarding entities, or metadata standards. It is fast and easy to deploy, as it takes advantage of the existing system like Wikidata and Wikipedia to access entities. 
 
\section{Related Work}
Various research works have addressed the issue of semantic table annotation. The most popular approaches which deal with the three above mentioned tasks rely on supervised learning setting, where candidate entities are selected by a classiﬁcation models \cite{ref_semtab19}. Such systems include (i) MTab \cite{ref_Mtab_19}, which combines a voting algorithm and the probability models to solve critical problems of the matching tasks, (ii) DAGOBAH \cite{ref_Dagobah_19} aiming at semantically annotating tables with Wikidata and DBpedia entities; more precisely it performs cell and column annotation and relationship identification, via a pipeline starting from a pre-processing step to enriching an existing knowledge graph using the table information; (iii) ADOG \cite{ref_Adog_19} is a system focused on leveraging the structure of a well-connected ontology graph which is extracted from different Knowledge Graphs to annotate structured or semi-structured data. In the latter approach, they combine in novel ways a set of existing technologies and algorithms  to automatically annotate structured and semi-structured records. It takes advantage of the native graph structure of ontologies to build a well-connected network on ontologies from different sources; (iv) Another example is described in \cite{ref_ELKG_19}. Its process is split into a Candidate Generation and a Candidate Selection phases. The former involves looking for relevant entities in knowledge bases, while the latter involves picking the top candidate using various techniques such as heuristics (the ‘TF-IDF’ approach) and machine learning (the Neural Network Ranking model).

In \cite{ref_articleZhang2017} the authors present TableMiner, a learning approach for a semantic table interpretation. This is essentially done by improving annotation accuracy by making innovative use of various types of contextual information both inside and outside tables as features for inference. Then, it reduces computational overheads by adopting an incremental, bootstrapping approach that starts by creating preliminary and partial annotations of a table using ‘sample’ data in the table, then using the outcome as ‘seed’ to guide interpretation of remaining contents. Following also a machine learning approach, \cite{ref_Takeoka2019} proposes Meimei. It combines a latent probabilistic model with multi-label classifiers. 

Other alternative approaches address only a single specific task. Thus, in the work of \cite{ref_Chen2019}, the authors focuses on column type prediction for tables without any metadata. Unlike traditional lexical matching-based methods, they follow a deep prediction model that can fully exploit tables’ contextual semantics, including table locality features learned by a Hybrid Neural Network (HNN), and inter-column semantics features learned by a knowledge base (KB) lookup and query answering algorithm. It exhibits good performance not only on individual table sets, but also when transferring from one table set to another. In the same vein,  a work conducted by \cite{ref_Ermilov2016} propose Taipan, which is able to recover the semantics of tables by identifying subject columns using a combination of structural and semantic features. 

From Web tables point of view, various works could be mentioned. Thus, in \cite{ref_Ritze2015} an iterative matching approach is described. It combines both schema and entity matching and is dedicated to matching large set of HTML tables with a cross-domain knowledge base. 
Similarly, TabEL  uses a collective classification technique to disambiguate all mentions in a given table \cite{ref_article_Bhagavatula2015}. Instead of using a strict mapping of types and relations into a reference knowledge base, TabEL uses soft constraints in its graphical model to sidestep errors introduced by an incomplete or noisy KB. It outperforms previous work on multiple datasets.

Overall, all the above mentioned approaches are based on a learning strategy. However, for the real-time application, there is a need to get the result as fast as possible. Another main limitation of these approaches is their reproducibility. Indeed, key explicit information concerning study parameters (particularly randomization control) and software environment are lacking.

The ultimate goal with AMALGAM, which could be categorized as a tabular data to KG matching system, is to provide a fast and efficient approach for tabular data to KG matching task.   

\section{The AMALGAM approach}

\begin{figure}[!ht]
\centering
\includegraphics[width=0.7\columnwidth]{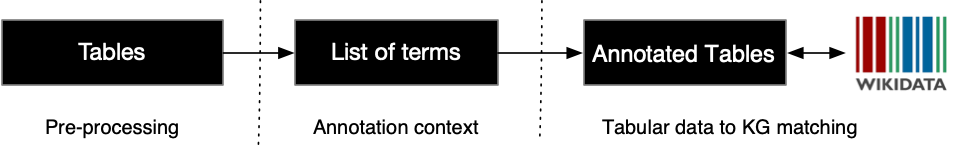}
\caption{Workflow of \texttt{AMALGAM}. 
} 
\label{fig1}
\end{figure}

\texttt{AMALGAM} is designed according to the workflow in Fig.~\ref{fig1}. There are three main phases which consist in, respectively, pre-processing, context annotation and tabular data to KG matching. The first two steps are identical for both CEA and CTA tasks.

\textbf{Tables Pre-Processing.} It is common to have missing values in datasets. Beside, the content of the table can have different types (string, date, float, etc.)
The aim of the pre-processing step is to ensure that loading table happens without any error. For instance, a textual encoding where some characters are loaded as noisy sequences or a text field with an unescaped delimiter will cause the considered record to have an extra column, etc. Loading incorrect encoding might strongly affect the lookup performance. To overcome this issue, \texttt{AMALGAM} relies on the Pandas library\footnote{https://pandas.pydata.org/} to fix all noisy textual data in the tables being processed.

\begin{figure}
\centering
\includegraphics[width=0.8\columnwidth]{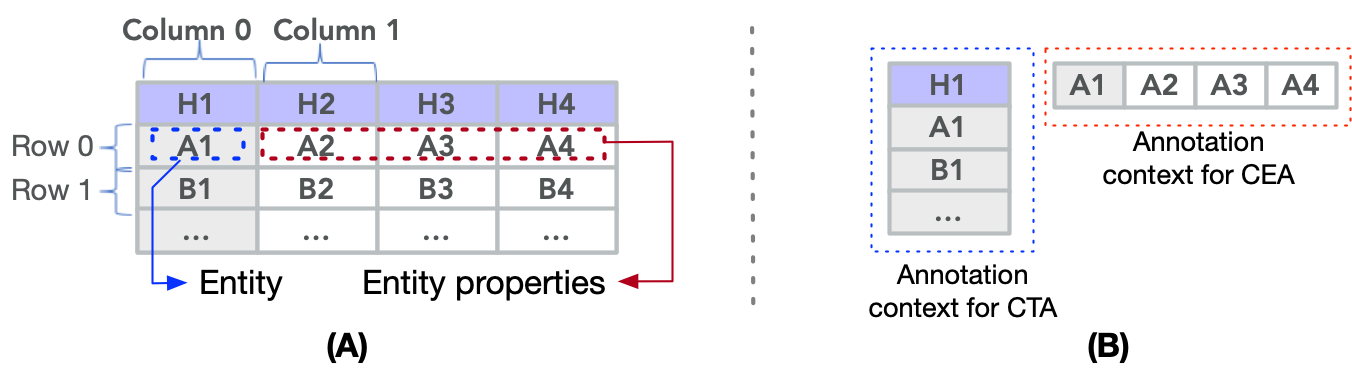}
\caption{Illustration of a table structure.} \label{fig2}
\end{figure}

\textbf{Annotation context.} We consider a table as a two-dimensional tabular structure (see Fig.~\ref{fig2}(A)) which is composed of an ordered set of x rows and y columns. Each intersection between a row and a column determines a cell $c_{ij}$ with the value $v_{ij}$ where $1 \leq i \leq x, 1 \leq j \leq y$. To identify the attribute label of a column also referred as header detection (\texttt{CTA} task), the idea consists in annotating all the items of the column using entity linking. Then, the attribute label is estimated using a random entity linking. The annotation context is represented by the list of items in the same column (see Fig.~\ref{fig2}(B)). For example, the context of the first column in the Fig.~\ref{fig2} is represented by the following items: [\textit{A1,B1,...,n}]. Following the same logic, we consider that all cells in the same row describe the same context. More specifically, the first cell of the row describes the entity and the following cells the associated properties. For instance, the context of the first row in the Fig.~\ref{fig2} is represented by the following list of items: [\textit{A1,A2,A3,A4}].

\begin{figure}[!ht]
\includegraphics[width=\textwidth]{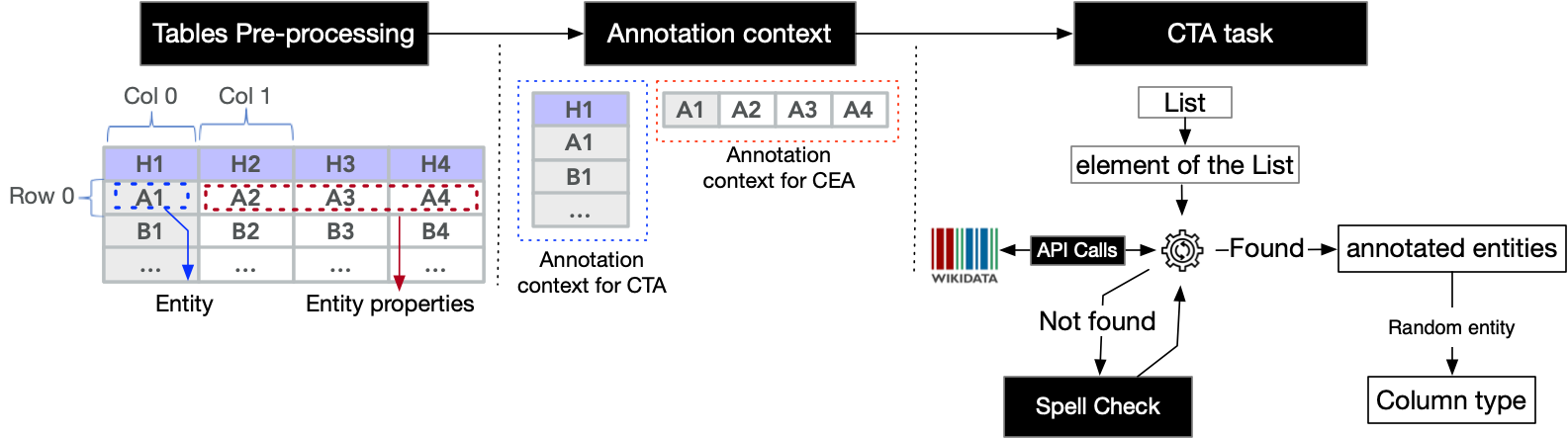}
\caption{Assigning a semantic type to a column (\texttt{CTA}).} \label{fig3}
\end{figure}

\textbf{Assigning a semantic type to a column (\texttt{CTA}).} The \texttt{CTA} task consists in assigning a Wikidata KG entity to a given column. It can be performed by exploiting the process described in Fig.~\ref{fig3}. The Wikidata KG allows to look up a Wikidata item by its title of its corresponding page on Wikipedia or other Wikimedia family sites using a dedicated API \footnote{https://www.wikidata.org/w/api.php}.  In our case, the main information needed from the entity is a list of the \textit{instances of (P31)}, \textit{subclass of (P279)} and \textit{part of (P361)} statements. To do so, a parser is developed to retrieve this information from the Wikidata built request. For example, "Grande Prairie" provides the following results: [list of towns in Alberta:Q15219391, village in Alberta:Q6644696, city in Alberta:Q55440238]. To achieve this, our methodology combines \textit{wbsearchentities} and \textit{parse} actions provided by the API.
It could be observed that in this task, there were many items that have not been annotated. This is because tables contain incorrectly spelled terms. Therefore, before implementing the other tasks, a spell check component is required.

As per the literature \cite{ref_article1}, spell-checker is a crucial language tool of natural language processing (\texttt{NLP}) which is used in applications like information extraction, proofreading, information retrieval, social media and search engines. In our case, we compared several approaches and libraries: Textblob\footnote{https://textblob.readthedocs.io/en/dev/}, Spark NLP\footnote{https://nlp.johnsnowlabs.com/}, Gurunudi\footnote{https://github.com/guruyuga/gurunudi}, Wikipedia api\footnote{https://wikipedia.readthedocs.io/en/latest/code.html}, Pyspellchecker\footnote{https://github.com/barrust/pyspellchecker}, Serpapi\footnote{https://serpapi.com/spell-check}. A comparison of these approaches could be found in table \ref{tab1}.

\begin{table}
\caption{Comparative of approaches and libraries related to spell-checking.}\label{tab1}
\footnotesize
\centering
\begin{tabular}{l|p{2.2cm}|p{7cm}}
\hline
\textbf{Name} &  \textbf{Category} & \textbf{Strengths/Limitations}\\
\hline
Textblob & NLP& Spelling correction, Easy-to-use \\
\hline
Spark NLP & NLP & Pre-trained, Text analysis\\
\hline
Gurunudi & NLP & Pre-trained, Text analysis, Easy-to-use\\
\hline
Wikipedia api & Search engines & Search/suggestion, Easy-to-use, Unlimited access  \\
\hline
Pyspellchecker & Spell checking & Simple algorithm, No pre-trained, Easy-to-use  \\
\hline
Serpapi & Search engines & Limited access for free\\
\hline
\end{tabular}
\end{table}

Our choice is oriented towards Gurunudi and the Wikidata API with a post-processing step consisting in validating the output using fuzzywuzzy\footnote{https://github.com/seatgeek/fuzzywuzzy} to keep only the results whose ratio is greater than the threshold of 90\%. For example, let's take the expression \textit{“St Peter's Seminarz”} after using the Wikidata API we get \textit{“St Peter's seminary”} and the ratio of fuzzy string matching is 95\%. 

We are now able to perform the \texttt{CTA} task. In the trivial case, the result of an item lookup is equal a single record. The best matching entity is chosen as a result. In the other cases, where the result is more than one, no annotation is produced for the \texttt{CTA} task. 
Finally, if there is no result after the lookup, another one is performed using the output of the spell check produced by the item. At the end of these lookups, the matched couple results are then stored in a nested dictionary [item:claims]. The most relevant candidate, counting the number of occurrences, is selected.



\begin{algorithm}[H]
\footnotesize 
\DontPrintSemicolon
\SetAlgoLined
\KwIn{\textit{$Table\ T$}}
\KwOut{$Annotated\ Table\ T'$}
\ForEach{$\textit{col i} \in \mathcal{T}{}$}{
    ${candidates\_col}_{} \gets \emptyset$\;
    \ForEach{$\textit{el} \in \textit{col}{}$}{
     ${label}_{} \gets \textit{el.value}$\;
     ${candidates}_{} \gets \textit{wd-lookup}\ (\textit{label})$\;
        \uIf{\textit{($candidates.size = 1 $)}}{
        $candidates\_col(k, candidates)$\;
        }\ElseIf{\textit{($candidates.size = 0 $)}}{
           ${new\_label} \gets \textit{spell-check}\ (\textit{label}) $\;
           ${candidates}_{} \gets \textit{wd-lookup}\ (\textit{new\_label})$\;
           \If{\textit{($candidates.size = 1 $)}}{
            $candidates\_col(k, candidates)$\;
           }
        }
     }
    $annotate(T'.col.i, getMostCommunClass(candidates\_col))$\;
 }
\caption{CTA task}
\end{algorithm}

\begin{figure}[!ht]
\includegraphics[width=\textwidth]{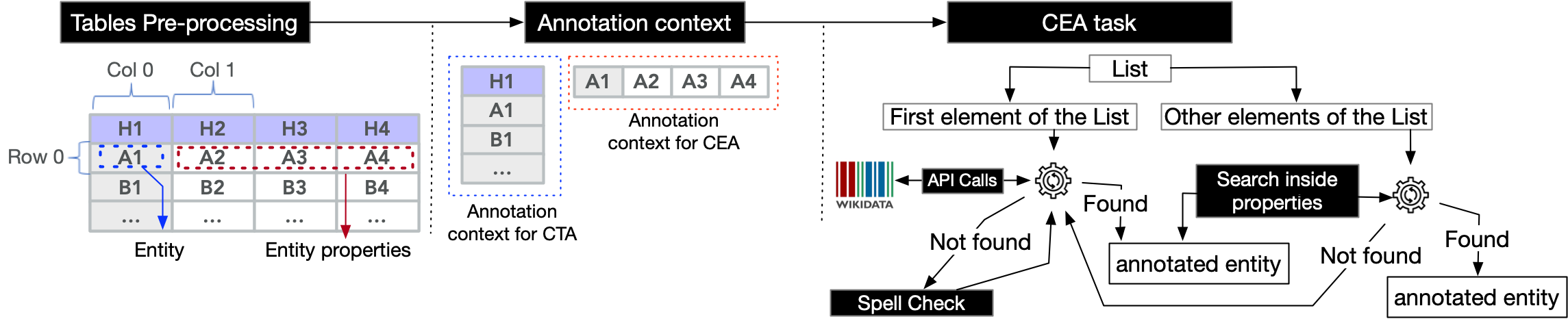}
\caption{Matching a cell to a KG entity (\texttt{CEA}).} \label{fig4}
\end{figure}

\textbf{Matching a cell to a KG entity (\texttt{CEA}).} The \texttt{CEA} task can be performed by exploiting the process described in Fig.~\ref{fig4}. Our approach reuse the process of the \texttt{CTA} task and made necessary adaptations. The first step is to get all the statements for the first item of the list context. The process is the same as \texttt{CTA}, the only difference is where result provides than one record. In this case, we create nested dictionary with all candidates. Then, to disambiguate the candidates entities, we use the concept of the column generated with the \texttt{CTA} task. Next, a lookup is performed by using the other items of the list context in the claims of the first item. If the item is found, it is selected as the target entity; if not, the lookup is performed with the item using the Wikidata API (if the result is empty, no annotation is produced). 

With this process, it is possible to reduce errors associated with the lookup. Let's take the value “650“ in row 0 of the table Fig.~\ref{fig4} for instance. If we lookup directly in Wikidata, we can get many results. However, if we check first in the statements of the first item of the list, “Grande Prairie“, it is more likely to successfully identify the item.


\begin{algorithm}[H]
\footnotesize 
\DontPrintSemicolon
\SetAlgoLined
\KwIn{\textit{$Table\ T$, $TColsContext$}}
\KwOut{$Annotated\ Table\ T'$}

\ForEach{$\textit{row i} \in \mathcal{T}{}$}{
    ${FirstEl\_properties}_{} \gets \emptyset$\;
    \ForEach{$\textit{el} \in \textit{row}{}$}{
     ${label}_{} \gets \textit{el.value}$\;
    \If{\textit{($el = 0 $)}}{
    ${FirstEl\_properties}_{} \gets GetProperties({label,ColsContext}_{})$ \;
    }
    \uIf{\textit{($\textit{Prop-lookup}\ (\textit{label}) \not= \emptyset $)}}{
        $annotate(T'.row.i.el, candidates.value)$\;
    }
    \Else{
        ${candidates}_{} \gets \textit{wd-lookup}\ (\textit{label,ColsContext})$ \;
        \uIf{\textit{($candidates.size = 1 $)}}{
        $annotate(T'.row.i.el, candidates.value)$ \;
        }\ElseIf{\textit{($candidates.size = 0 $)}}{
           ${new\_label} \gets \textit{spell-check}\ (\textit{label}) $\;
           ${candidates}_{} \gets \textit{wd-lookup}\ (\textit{new\_label,ColsContext})$ \;
           \If{\textit{($candidates.size = 1 $)}}{
            $annotate(T'.row.i.el, candidates.value)$\;
           }
        }
    }

     }
 }
\caption{Algorithm of CEA processing task}
\end{algorithm}


\section{Experimental Results}
The evaluation of AMALGAM is done in the context of the SemTab 2020 challenge\footnote{http://www.cs.ox.ac.uk/isg/challenges/sem-tab/2020/index.html}. This challenge is subdivided into 4 successive rounds containing respectively 34294, 12173, 62614 and 22387 CSV tables to annotate.
For example, Table~\ref{tab2}, lists all Alberta towns with additional information such as the country and the elevation above sea level. The evaluation metrics are respectively the F1 score and the Precision \cite{ref_semtab_iswc2020}.
\begin{table}
\caption{List of Alberta towns, extracted from SemTab Round 1.}\label{tab2}
\centering
\footnotesize
\begin{tabular}{l|l|l|l|l|l}
\hline
col0 &  col1 & col2 &col3&col4 & col5\\
\hline
Grande Prairie &city in Alberta&canada&Sexsmith&650&Alberta\\
\hline
Sundre	&town in Alberta&canada&Mountain View County&1093&Alberta\\
\hline
Peace River & town in clberta&Canada & Northern Sunrise County & 330 & Alberta\\
\hline
Vegreville&town in Alberta&canada&Mundare&635&Alberta\\
\hline
\end{tabular}
\end{table}

Tables 3, 4, 5 and 6 report the evaluation of \texttt{CTA} and \texttt{CEA} respectively for round 1, 2, 3 and 4. Thus, it could be observed that \texttt{AMALGAM} handles properly the two tasks, in particular in the \texttt{CEA} task. Regarding the \texttt{CTA} task, these results can be explained with a new revision of Wikidata created in the item revision history and there are possibly spelling errors in the contents of the tables. For instance, "rural district of Lower Saxony" became "district of Lower Saxony" after the 16th April 2020 revision. A possible solution to this issue is to retrieve the history of the different revisions, by parsing Wikidata data history dumps, to use them in the lookup. This is a possible extension to this work. Another observed issue is that spelling errors impacts greatly the lookup eﬃciency.

\begin{table}[!htb]
    \begin{minipage}{.5\linewidth}
      \caption{Results of Round 1.}
      \centering
        \begin{tabular}{|l|l|l|}
        \hline
            TASK & F1 Score & Precision \\
        \hline
        CTA & 0.724 & 0.727\\
        CEA & 0.913 & 0.914	\\
        \hline
        \end{tabular}
    \end{minipage}%
    \begin{minipage}{.5\linewidth}
      \centering
        \caption{Results of Round 2.}
        \begin{tabular}{|l|l|l|}
        \hline
            TASK & F1 Score & Precision \\
        \hline
        CTA &0.926 & 0.928\\
        CEA &0.921 & 0.927\\
        \hline
        \end{tabular}
    \end{minipage} 
\end{table}

\begin{table}[!htb]
    \begin{minipage}{.5\linewidth}
      \caption{Results of Round 3.}
      \centering
        \begin{tabular}{|l|l|l|}
        \hline
            TASK & F1 Score & Precision \\
        \hline
        CTA & 0.869 & 0.873\\
        CEA & 0.877 & 0.892 \\
        \hline
        \end{tabular}
    \end{minipage}%
    \begin{minipage}{.5\linewidth}
      \centering
        \caption{Results of Round 4.}
        \begin{tabular}{|l|l|l|}
        \hline
            TASK & F1 Score & Precision \\
        \hline
        CTA &0.858 & 0.861\\
        CEA &0.892 & 0.914\\
        \hline
        \end{tabular}
    \end{minipage} 
\end{table}

From the round 1 experience, we specifically focused on the spell check process of items to improve the results of the \texttt{CEA} and \texttt{CTA} tasks in round 2. Two API services, from Wikipedia and Gurunudi (presented in Sect. 3.) respectively were used for spelling correction. According to the results in Table 4, both F1-Score and Precision have been improved. From these rounds, we observed that term with a single word is often ambiguous as it may refer to more than one entity. In Wikidata, there is only one article (one entry) for each concept. However, there can be many equivalent titles for a concept due to the existence of synonyms, etc.  These synonymy and ambiguity issues make it difficult to match the correct item. For example, the term “Paris” may refer to various concepts such as “the capital and largest city of France“, “son of Priam, king of Troy“, “county seat of Lamar County, Texas, United States“. This leads us to introduce a disambiguation process during rounds 3 and 4. For these two last rounds, we have updated the annotation algorithm by integrating the concept of the column obtained during the \texttt{CTA} task in the linking phase. We showed that the two tasks can be performed relatively successfully with \texttt{AMALGAM}, achieving higher than 0.86 in precision. However, the automated disambiguation process of items proved to be a more challenging task.

\section{Conclusion and Future Works}
In this paper, we described \texttt{AMALGAM}, a matching approach to enabling tabular datasets to be FAIR compliant by making them explicit thanks to their annotation using a knowledge graph, in our case Wikidata. Its advantage is that it allows to perform both \texttt{CTA} and \texttt{CEA} tasks in a timely manner. These tasks can be accomplished through the combination of lookup services and a spell check techniques quickly. The results achieved in the context of the SemTab 2020 challenge show that it handles table annotation tasks with a promising performance. Our findings suggest that the matching process is very sensitive to errors in spelling. Thus, as of future work, an improved spell checking techniques will be investigated. Further, to process such errors the contextual based spell-checkers are needed. Often the string is very close in spelling, but context could help reveal which word makes the most sense. Moreover, the approach will be improved through finding a trade-off between effectiveness and efficiency.
%
%

\end{document}